\documentclass[12pt]{article}

\usepackage{graphicx}
\begin{document}

\begin{center}
{\bf Universe inflation and nonlinear electrodynamics}\\
\vspace{5mm} S. I. Kruglov
\footnote{E-mail: serguei.krouglov@utoronto.ca}
\underline{}
\vspace{3mm}

\textit{Department of Physics, University of Toronto, \\60 St. Georges St.,
Toronto, ON M5S 1A7, Canada\\
Canadian Quantum Research Center, \\
204-3002 32 Ave., Vernon, BC V1T 2L7, Canada} \\
\vspace{5mm}
\end{center}

\begin{abstract}

We analyse the universe inflation when the source of gravity is electromagnetic fields obeying nonlinear electrodynamics with two parameters and without singularities. The cosmology of the universe with stochastic magnetic fields is considered. The condition for the universe inflation is obtained. It is demonstrated that singularities of the energy density and pressure are absent as the scale factor approaches to zero. When the scale factor goes to infinity one has equation of state for ultra-relativistic case. The curvature invariants do not possess singularities. The evolution of universe is described showing that at large time the scale factor corresponds to the radiation era. The duration of universe inflation is analysed. We study the classical stability and causality by computing the speed of the sound. Cosmological parameters such as the spectral index $n_s$, the tensor-to-scalar ratio $r$ and the running of the spectral index $\alpha_s$ are evaluated.

\end{abstract}

\section{Introduction}

Universe inflation may be justified by the modification of general relativity ($F(R)$-gravity) \cite{Starobinsky,Capozziello,Nojiri}, insertion of cosmological constant $\Lambda$ in Einstein--Hilbert action or by introduction of a scalar field (quintessence) \cite{Linde}. Another way to explain universe acceleration is to use the source of Einstein' gravity in the form of nonlinear electrodynamics (NED). First NED was proposed by Born and Infeld \cite{Born} to smooth a singularity of point-like charges and to have self-energy finite.
It is worth noting that at strong electromagnetic fields classical Maxwell's electrodynamics becomes NED due to quantum corrections \cite{Heisenberg,Schwinger,Adler}. Thus, we imply that in early time of the universe evolution electromagnetic fields were very strong and linear Maxwell's electrodynamics should be replaced by NED. Inflation can be described by Einstein' gravity coupled to some NED \cite{Camara,Elizalde,Novello3,Novello,Novello1,Vollick,Salcedo,Kruglov1,Kruglov0,Kruglov2,Kruglov3,Kruglov4}. We explore here NED with two parameters \cite{Kruglov5} which includes rational NED \cite{Kruglov1,Kruglov2} for some parameter and for weak fields it becomes Maxwell's electrodynamics.
We show that in the framework of our model the universe inflation takes place for the stochastic magnetic background field.

It was shown that the stochastic fluctuations of the electromagnetic field are present in a relativistic electron-positron plasma and, therefore, plasma fluctuations can generate a stochastic magnetic field  \cite{Lemoine,Lemoine1}. Thermal fluctuations in the pre-recombination plasma could be the source of a primordial magnetic field and magnetic fluctuations may be sustained by plasma before the epoch of Big Bang nucleosynthesis. Probably, there were strong low-frequency random magnetic fields in the early stage of the radiation-dominated era.
In our galaxy and other spiral galaxies magnetic fields of the order of $B = 10^{-6}$ G are present on scales of several Kpc  \cite{Kronberg}.
The fields $B = 10^{-6}$ G can have the primordial origin and are explained with the help of the galactic dynamo theory
This theory explains a mechanism transferring angular momentum energy into magnetic energy and needs the existence of weak seed fields of the order of $B= 10^{-19}$ G at the epoch of the galaxy formation. The source of seed magnetic fields could be thermal fluctuations in the primordial plasma. Magnetic energy over larger scales can be due to long wavelength fluctuations which reconnect and redistribute magnetic fields \cite{Gaensler}. As the electric field is screened by the charged primordial plasma we consider the background magnetic field \cite{Lemoine1}.
We imply that directional effects are not present ($\langle\textbf{B}\rangle = 0$) according to the standard cosmological model.

In section 2 we consider Einstein's gravity coupled to NED with dimensional parameter $\beta$ and dimensionless parameter $\gamma$. The source of gravity is a stochastic magnetic field. The range of magnetic fields, when an acceleration of universe occurs, is obtained. We show that singularities of the energy density and pressure are absent at $\gamma \geq 1$. It is demonstrated that, when the scale factor approaches infinity, one has equation of state (EoS) for ultra-relativistic case. Singularities of curvature invariants
are absent for $\gamma\geq 1$.  The evolution of the universe is investigated in section 3. We find the dependence of the scale factor on the time and show that when time goes to infinity one has the radiation era. In subsection 3.1, the deceleration parameter as a function of the scale factor is computed showing the evolution of universe. We analyse the amount of the inflation by calculating e-foldings $N$. For some parameters the reasonable e-foldings $N\approx70$ is realised. We found the duration of the universe inflation depending on parameters beta and gamma.
In subsection 3.2 the causality and unitarity principles are studied by computing the speed of sound. In section 4 the cosmological parameters, the spectral index $n_s$, the tensor-to-scalar ratio $r$, and the running of the spectral index $\alpha_s$, are evaluated. We show that they are in approximate agreement with the PLANK and WMAP data. Section 5 is a Summary.

Units with $c=\hbar=1$ are used and the metric signature is $\eta=\mbox{diag}(-,+,+,+)$.

\section{Cosmology with a stochastic magnetic field of background}

The Einstein--Hilbert action coupled to the matter source is given by
\begin{equation}
S=\int d^4x\sqrt{-g}\left[\frac{1}{2\kappa^2}R+ {\cal L}\right],
\label{1}
\end{equation}
where $R$ is the Ricci scalar, $\kappa^2=8\pi G$ and $G$ is Newton's constant with the dimension $[L^2]$. We consider the source of gravitational field in the form of NED with the Lagrangian \cite{Kruglov5}
\begin{equation}
{\cal L}=-\frac{{\cal F}}{1+\epsilon(2\epsilon\beta {\cal F})^\gamma}.
\label{2}
\end{equation}
Here, ${\cal F}=F^{\mu\nu}F_{\mu\nu}/4=(B^2-E^2)/2$ is the field invariant and $E$ and $B$ are the electric and magnetic fields, correspondingly. We introduce the parameter $\epsilon=\pm 1$, dimensional  parameter $\beta>0$ with the dimension $[L^4]$, and dimensionless parameter $\gamma>0$. If $B>E$ one uses $\epsilon=1$ and for $B<E$ we should set $\epsilon=-1$ to have the real Lagrangian. It is worth noting that as $\beta {\cal F}\rightarrow 0$ Lagrangian (2) approaches to Maxwell's Lagrangian. It was shown in Ref. \cite{Kruglov6} that there are no singularities in the electric field of point-like charges and electric self-energy in the model with the Lagrangian (2) for $\epsilon=-1$. Here, we consider inflation of the universe filled by
stochastic magnetic fields ($\epsilon= 1$). After varying action (1), one obtains the Einstein and electromagnetic field equations
\begin{equation}
R_{\mu\nu}-\frac{1}{2}g_{\mu\nu}R=\kappa^2T_{\mu\nu},
\label{3}
\end{equation}
\begin{equation}
\nabla_\mu({\cal L}_{\cal F}F^{\mu\nu})=0,
\label{4}
\end{equation}
where
\begin{equation}
{\cal L}_{\cal F}=\frac{\partial {\cal L}}{\partial {\cal F}}=-\frac{1-\epsilon(\gamma-1)(2\epsilon\beta {\cal F})^\gamma}{(1+\epsilon(2\epsilon\beta {\cal F})^\gamma)^2},
\label{5}
\end{equation}
and the stress-energy tensor is given by
\begin{equation}
 T_{\mu\nu }=-F_{\mu\rho }F_{\nu }^{~\rho }\mathcal{L}_{\mathcal{F}}-g_{\mu \nu }\mathcal{L}\left( \mathcal{F}\right).
\label{6}
\end{equation}
We use the line element of homogeneous and isotropic cosmological space-time as follows:
\begin{equation}
ds^2=-dt^2+a(t)^2\left(dx^2+dy^2+dz^2\right),
\label{7}
\end{equation}
with a scale factor $a(t)$. Let us consider the cosmic background in the form of stochastic magnetic fields. As the electric field is zero we use $\epsilon=1$.  We have the isotropy of the Friedman--Robertson--Walker space-time after averaging the magnetic fields \cite{Tolman}.
It is implied that the wavelength of electromagnetic waves is smaller that the curvature. Therefore, one should impose equations as
\begin{equation}
\langle\textbf{B}\rangle=0,~~~~\langle B_iB_j\rangle=\frac{1}{3}B^2g_{ij},
\label{8}
\end{equation}
with the brackets denoting an average over a volume. For simplicity we will omit the brackets in the following. Note that the NED stress-energy tensor can be represented in the form of a perfect fluid \cite{Novello1}. The Friedmann's equation for three dimensional flat universe is
\begin{equation}
3\frac{\ddot{a}}{a}=-\frac{\kappa^2}{2}\left(\rho+3p\right),
\label{9}
\end{equation}
where $\ddot{a}=\partial^2a/dt^2$.
When $\rho + 3p < 0$ the universe acceleration takes place. Making use of Eq. (6) we obtain
\[
\rho=-{\cal L}=\frac{B^2}{2[1+(\beta B^2)^\gamma]},
\]
\begin{equation}
p={\cal L}-\frac{2B^2}{3}{\cal L}_{\cal F}=\frac{B^2[1+(1-4\gamma)(\beta B^2)^\gamma]}{6[1+(\beta B^2)^\gamma]^2}.
\label{10}
\end{equation}
By virtue of Eq. (10) one finds
\begin{equation}
\rho+3p=\frac{B^2[1+(1-2\gamma)(\beta B^2)^\gamma]}{[1+(\beta B^2)^\gamma]^2}.
\label{11}
\end{equation}
From Eq. (11) and the condition $\rho + 3p < 0$ for the acceleration of the universe, we obtain
\begin{equation}
\sqrt{\beta}B>\frac{1}{(2\gamma-1)^{1/(2\gamma)}}.
\label{12}
\end{equation}
with $\gamma>1/2$. The plot of the function $\sqrt{\beta}B$ versus $\gamma$ is given in Fig. 1.
\begin{figure}[h]
\includegraphics[height=4.0in,width=4.0in]{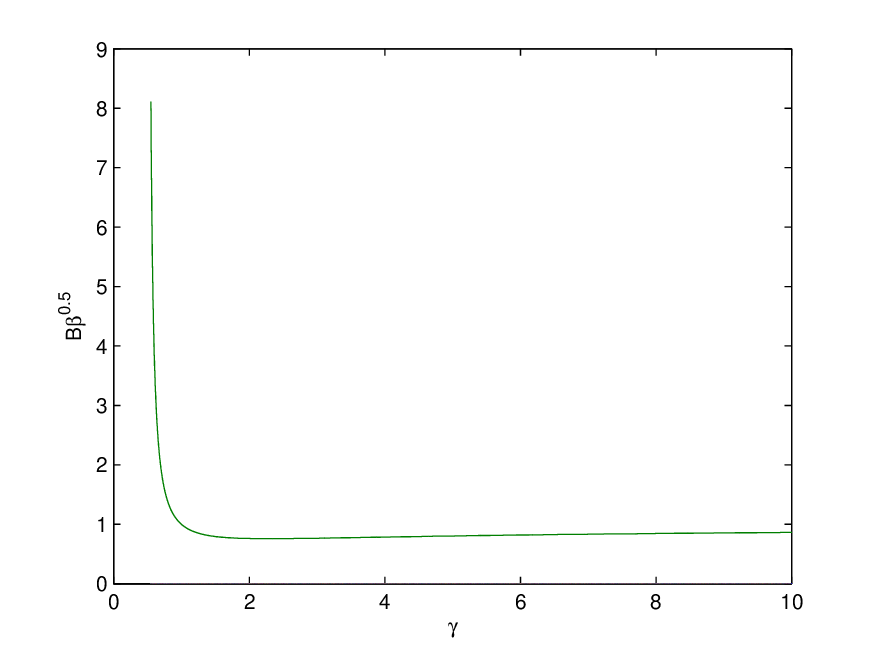}
\caption{\label{fig.1}The function  $\sqrt{\beta}B$ vs. $\gamma$. The minimum of the function $\sqrt{\beta}B$ occurs at $\approx 2.296$. At $\gamma\rightarrow 1/2$, $\sqrt{\beta}B\rightarrow\infty$ and as $\gamma\rightarrow\infty$, we have $\sqrt{\beta}B\rightarrow 1$. }.
\end{figure}
The minimum value of $\sqrt{\beta}B$ takes place at $\gamma=0.5 \exp\left(W(1/e)+1\right)\approx 2.296$, where $W(z)$ is the Lambert function.
Thus, the strong magnetic fields drives the universe inflation. The stress-energy tensor conservation, $\nabla^\mu T_{\mu\nu}=0$, leads to the equation
\begin{equation}
\dot{\rho}+3\frac{\dot{a}}{a}\left(\rho+p\right)=0.
\label{13}
\end{equation}
Making use of Eq. (10), one finds
\begin{equation}
\rho+p=\frac{2B^2[1-(\gamma-1)(\beta B^2)^\gamma]}{3\left[1+(\beta B^2)^\gamma\right]^2}.
\label{14}
\end{equation}
After integrating Eq. (13), with the help of Eq. (14), we obtain
\begin{equation}
B(t)=\frac{B_0}{a(t)^2},
\label{15}
\end{equation}
were $B_0$ is the magnetic field corresponding to the value $a(t)=1$. It is worth mentioning that Eq. (15) holds for any Lagrangians \cite{Novello1}. Because the scale factor increases during the inflation, the magnetic field decreases. By virtue of Eqs. (10) and (15) one finds
\[
\lim_{a(t)\rightarrow \infty}\rho(t)=\lim_{a(t)\rightarrow \infty}p(t)=0,
\]
\[
\lim_{a(t)\rightarrow 0}\rho(t)=\lim_{a(t)\rightarrow 0}p(t)=\infty~~~~~~~\gamma<1,
\]
\[
\lim_{a(t)\rightarrow 0}\rho(t)=\lim_{a(t)\rightarrow 0}p(t)=0~~~~~~~\gamma>1,
\]
\begin{equation}
\lim_{a(t)\rightarrow 0}\rho(t)= - \lim_{a(t)\rightarrow 0}p(t)=\frac{1}{2\beta}~~~~~~\gamma=1.
\label{16}
\end{equation}
As a result, singularities of the energy density and pressure as $a(t)\rightarrow 0$ are absent at $\gamma\geq 1$. Equation (16) shows that at the beginning of the universe evolution ($a \approx 0$), the model gives $\rho=-p$ at $\gamma=1$ for the rational NED, corresponding to de Sitter space-time, i.e.  we have the same property as in the $\Lambda CDM$ model. Making use of Eq. (10) we obtain EoS
\begin{equation}
w=p(t)/\rho(t)=\frac{1+(1-4\gamma)(\beta B^2)^\gamma}{3[1+(\beta B^2)^\gamma]}.
\label{17}
\end{equation}
The function $w$ versus $(\beta B^2)^\gamma$ is depicted in Fig. 2 for $\gamma=0.75,~ 1,~ 1.5$.
\begin{figure}[h]
\includegraphics[height=4.0in,width=4.0in]{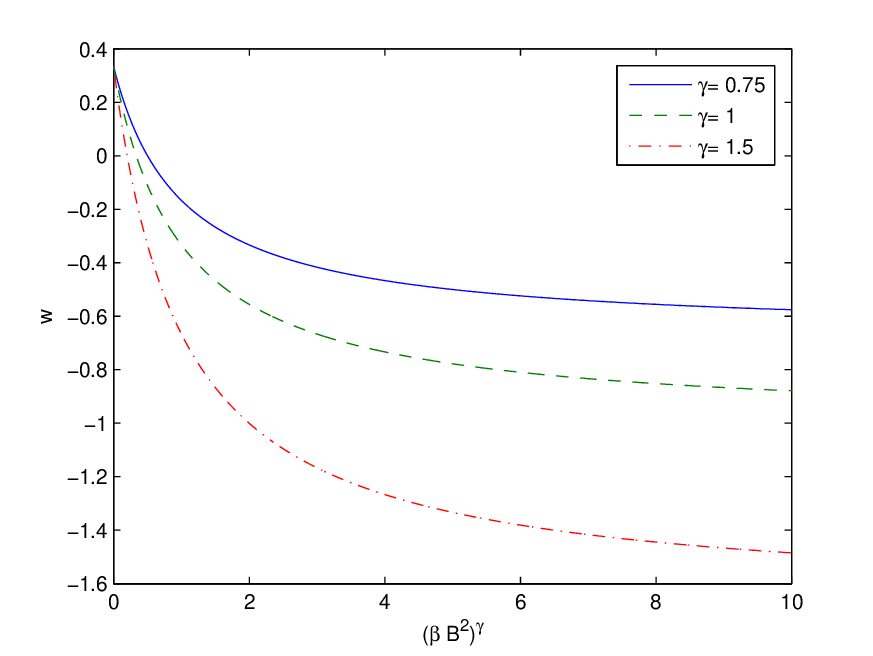}
\caption{\label{fig.2}The function  $w$ vs. $(\beta B^2)^{\gamma}$ for $\gamma=0.75, 1, 1.5$.}
\end{figure}
Making use of Eqs. (17)one finds
\begin{equation}
\lim_{B\rightarrow 0} w=\frac{1}{3}.
\label{18}
\end{equation}
As $a(t)\rightarrow \infty$ ($B\rightarrow 0$) one has the EoS for ultra-relativistic case \cite{Landau}.
For $\gamma>1$ and $(\beta B^2)^\gamma=1/(\gamma-1)$, one has de Sitter space-time, $w=-1$. Thus, for an example, at $\gamma=1.5$ and
$(\beta B^2)^{1.5}=2$ one has $w=-1$ (see Fig. 2).
From Eqs. (3) and (6), we obtain the Ricci scalar
\begin{equation}
R=\kappa^2T_\mu^{~\mu}=\frac{2\kappa^2\gamma B^2(\beta B^2)^\gamma}{[1+(\beta B^2)^\gamma]^2}=\kappa^2\left[\rho(t)-3p(t)\right].
\label{19}
\end{equation}
The function $\beta R/\kappa^2$ versus $(\beta B^2)^{\gamma}$ is depicted in Fig. 3.
\begin{figure}[h]
\includegraphics[height=4.0in,width=4.0in]{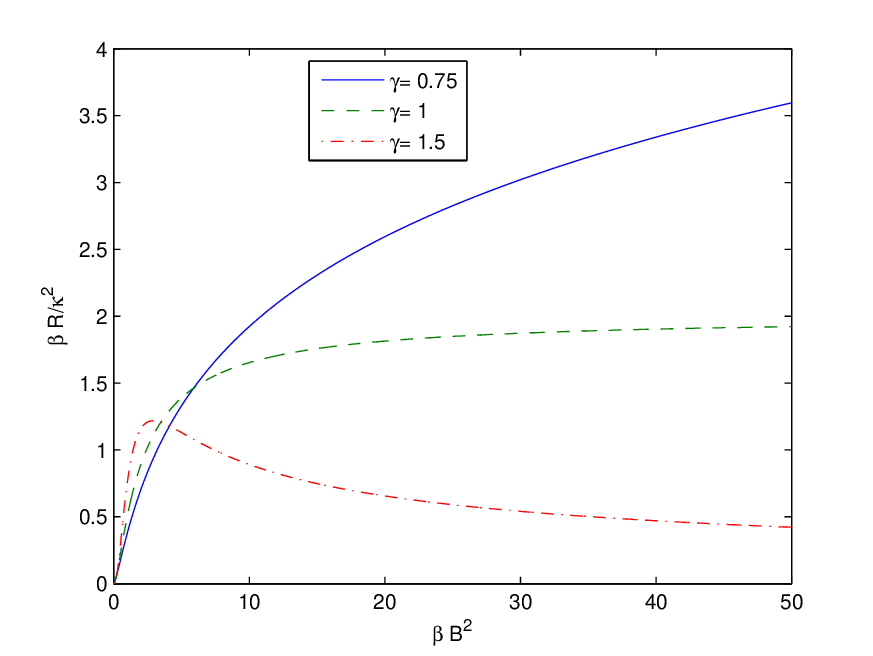}
\caption{\label{fig.3}The function  $\beta R/\kappa^2$ vs. $\beta B^2$ for $\gamma=0.75,~ 1,~ 1.5$. For $\gamma\geq1$ the singularity of the Ricci scalar is absent. }
\end{figure}
Making use of Eq. (19) we find
\[
\lim_{B\rightarrow 0}R(t)=0,
\]
\[
\lim_{B\rightarrow \infty}R(t)=0~~~\mbox{at}~~\gamma>1,
\]
\begin{equation}
\lim_{B\rightarrow \infty}R(t)=\frac{2\kappa^2}{\beta}~~~\mbox{at}~~\gamma=1.
\label{20}
\end{equation}
Note that when the scale factor $a(t)\rightarrow 0$ one has $B\rightarrow \infty$ and the singularity of the Ricci scalar is absent for $\gamma\geq 1$. The Ricci tensor squared $R_{\mu\nu}R^{\mu\nu}$ and the Kretschmann scalar $R_{\mu\nu\alpha\beta}R^{\mu\nu\alpha\beta}$ may be expressed as linear combinations of $\kappa^4\rho^2$, $\kappa^4\rho p$, and $\kappa^4p^2$ \cite{Kruglov1} and, according to Eq. (16), they are finite as $a(t)\rightarrow 0$ and $a(t)\rightarrow \infty$ for $\gamma\geq1$.
As $t\rightarrow\infty$ the scale factor increases and spacetime approaches to the Minkowski space-time. From Eqs. (12) and (15) we find that the universe accelerates at $a(t)<\beta^{1/4}\sqrt{B_0}(2\gamma-1)^{1/(4\gamma)}$ and the universe inflation occurs.

\section{ Evolution of the universe}

We will use the Friedmann equation for three dimensional flat universe to describe evolution of the universe which is given by
\begin{equation}
\left(\frac{\dot{a}}{a}\right)^2=\frac{\kappa^2\rho}{3}.
\label{21}
\end{equation}
Making use of Eqs. (10) and (21), one obtains
\begin{equation}
\dot{a} =\frac{\kappa B_0a^{2\gamma-1}}{\sqrt{6}\sqrt{a^{4\gamma}+(\beta B_0^2)^{\gamma}}}.
\label{22}
\end{equation}
With the help  of the unitless variables $x\equiv a/(\beta^{1/4}\sqrt{B_0})$, $y\equiv\sqrt{6\beta}\dot{x}/\kappa$ Eq. (22) becomes
\begin{equation}
y =\frac{ x^{2\gamma-1}}{\sqrt{x^{4\gamma}+1}}.
\label{23}
\end{equation}
The function $y$ versus $x$ is depicted in Fig. 4.
\begin{figure}[h]
\includegraphics[height=4.0in,width=4.0in]{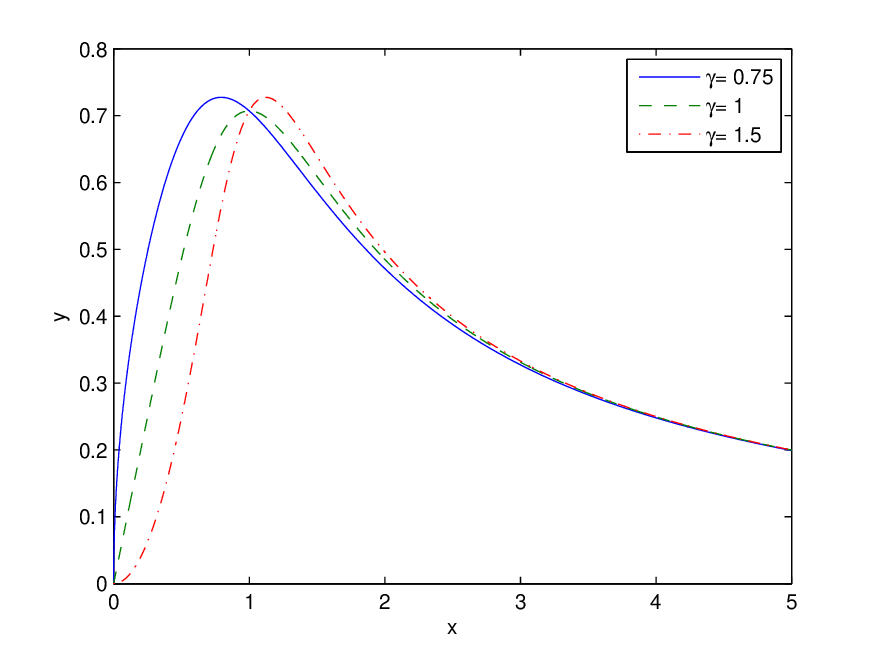}
\caption{\label{fig.4}The function $y=\sqrt{6\beta}\dot{x}/\kappa$ vs. $x=a/(\beta^{1/4}\sqrt{B_0})$ for $\gamma=0.75,~1,~1.5$.}
\end{figure}
According to Fig. 4, the inflation lasts from Big Bang till the graceful exit, $x_{end}=(2\gamma-1)^{1/(4\gamma)}$, and then the universe decelerates. After integrating Eq. (22) we obtain
\begin{equation}
\int_{a(t_{in})}^{a(t_{end})} \frac{\sqrt{a^{4\gamma}+(\beta B_0^2)^\gamma}}{a^{2\gamma-1}}da =\frac{\kappa B_0}{\sqrt{6}}\int_{t_{in}}^{t_{end}} dt.
\label{24}
\end{equation}
From Eq. (24) we arrive at the equation
\[
\frac{a(t_{end})^{2(1-\gamma)}(\beta B_0^2)^{\gamma/2}}{2(1-\gamma)}F\left(-\frac{1}{2},\frac{1}{2\gamma}-\frac{1}{2};\frac{1}{2\gamma}+\frac{1}{2};-\frac{a(t_{end})^{4\gamma}}{(\beta B_0^2)^\gamma}\right)
\]
\begin{equation}
-\frac{a(t_{in})^{2(1-\gamma)}(\beta B_0^2)^{\gamma/2}}{2(1-\gamma)}F\left(-\frac{1}{2},\frac{1}{2\gamma}-\frac{1}{2};\frac{1}{2\gamma}+\frac{1}{2};-\frac{a(t_{in})^{4\gamma}}{(\beta B_0^2)^\gamma}\right)=\frac{\kappa B_0}{\sqrt{6}}\Delta t,
\label{25}
\end{equation}
where $F(a,b;c;z)$ is the hypergeometric function and $\triangle t$ is the duration of the inflation.  It is worth noting that the hypergeometric function entering Eq. (25) has the property $c-b=1$, and therefore, it can be expressed through the incomplete $B$-function by the relation $B_z(p,q)=p^{-1}z^pF(1-q,p;p+1;z)$ \cite{Bateman}. Equation (25) allows us to study the evolution of the universe inflation.

\subsection{Universe inflation}

The expansion of the universe can be described  by the deceleration parameter. From Eqs. (9), (10), (15) and (21) we obtain the deceleration parameter
\begin{equation}
q=-\frac{\ddot{a}a}{(\dot{a})^2}=\frac{x^{4\gamma}+1-2\gamma}{x^{4\gamma}+1}.
\label{26}
\end{equation}
\begin{figure}[h]
\includegraphics[height=4.0in,width=4.0in]{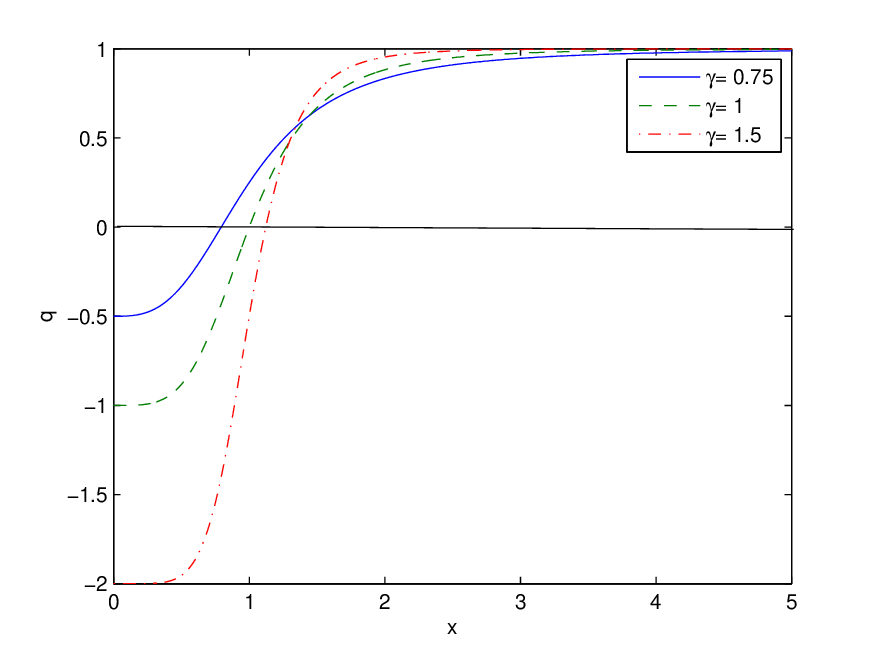}
\caption{\label{fig.5}The function $q$ vs. $x=a/(\beta B_0^2)^{1/4}$.}
\end{figure}
The plot of the deceleration parameter $q$ versus $x=a/(\beta B_0^2)^{1/4}$ is depicted in Fig. 5. The inflation, $q<0$, takes place till the graceful exit $q=0$ ($x_{end}=(2\gamma-1)^{1/(4\gamma)}$). When $x=(2\gamma-1)^{1/(4\gamma)}$ the acceleration stops, the deceleration parameter is zero, and then one has the deceleration phase ($q>0$). Singularities at the early epoch are absent.

Now, we evaluate the amount of the inflation with the help of e-foldings \cite{Liddle}
\begin{equation}
N=\ln\frac{a(t_{end})}{a(t_{in})},
\label{27}
\end{equation}
where $t_{end}$ corresponds to the final time of the inflation and $t_{in}$ is an initial time. The graceful exit point is $x_{end}= (2\gamma-1)^{1/(4\gamma)}$ and one finds $a(t_{end})= (2\gamma-1)^{1/(4\gamma)} b$ ($b= \beta^{1/4}\sqrt{B_0}$).
It is known that the horizon and flatness problems can be solved when e-foldings $N\approx 70$ \cite{Liddle}. From Eq. (27) we find the scale factor corresponding to the inflation initial time
\begin{equation}
a(t_{in})=\frac{(2\gamma-1)^{1/(4\gamma)}b}{\exp(70)}\approx (2\gamma-1)^{1/(4\gamma)}\times 10^{-31}b.
\label{28}
\end{equation}
One can analyze different scenarios of the universe inflation by varying $\gamma$, $b$, and $N$. This model possesses phases of the universe inflation, the graceful exit and deceleration.
Making use of Eq. (25) and condition $a(t_{end})\gg a(t_{in})$ ($a(t_{end})= 10^{31}a(t_{in})$), we obtain
\begin{equation}
\triangle t=\frac{\sqrt{3}a(t_{end})^{2(1-\gamma)}\beta^{\gamma/2}B_0^{\gamma-1}}{\sqrt{2}(1-\gamma)\kappa}
F\left(-\frac{1}{2},\frac{1}{2\gamma}-\frac{1}{2};\frac{1}{2\gamma}+\frac{1}{2};-\frac{a(t_{end})^{4\gamma}}{(\beta B_0^2)^\gamma}\right).
\label{29}
\end{equation}
To study the scale factor at $a(t_{end})\gg  1$, we use the Pfaff transformation \cite{Bateman}
\begin{equation}
F(a,b;c;z)=(1-z)^{-a}F\left(a,c-b;c;\frac{z}{z-1}\right).
\label{30}
\end{equation}
By virtue of Eq. (30), at $a(t_{end})\gg 1$, we obtain
\[
F\left(-\frac{1}{2},\frac{1}{2\gamma}-\frac{1}{2};\frac{1}{2\gamma}+\frac{1}{2};-\frac{a(t_{end})^{4\gamma}}{(\beta B_0^2)^\gamma}\right)\approx \sqrt{1+\frac{a(t_{end})^{4\gamma}}{(\beta B_0^2)^\gamma}}F\left(-\frac{1}{2},1;\frac{\gamma+1}{2\gamma};1\right)
\]
\begin{equation}
=\sqrt{1+\frac{a(t_{end})^{4\gamma}}{(\beta B_0^2)^\gamma}}
\frac{\Gamma\left(\frac{\gamma+1}{2\gamma}\right)\Gamma\left(\frac{1}{2\gamma}\right)}{\Gamma\left(\frac{2\gamma+1}{2\gamma}\right)
\Gamma\left(\frac{1-\gamma}{2\gamma}\right)},
\label{31}
\end{equation}
where $\Gamma(x)$ is Gamma-function. From Eqs. (29) and (31), at $a(t_{end})\gg 1$, we obtain  in the leading order the equation as follows:
\begin{equation}
\frac{a(t_{end})^2}{2(1-\gamma)}\frac{\Gamma\left(\frac{\gamma+1}{2\gamma}\right)\Gamma\left(\frac{1}{2\gamma}\right)}{\Gamma\left(\frac{2\gamma+1}{2\gamma}\right)
\Gamma\left(\frac{1-\gamma}{2\gamma}\right)}=\frac{\kappa B_0}{\sqrt{6}}\Delta t.
\label{32}
\end{equation}
Making use of Eq. (32), at $a(t_{end})\gg 1$,  we arrive at
\begin{equation}
a(t_{end})== \sqrt{\frac{2\kappa B_0(1-\gamma)}{\sqrt{6}}\frac{\Gamma\left(\frac{2\gamma+1}{2\gamma}\right)\Gamma\left(\frac{1-\gamma}{2\gamma}\right)}
{\Gamma\left(\frac{\gamma+1}{2\gamma}\right)\Gamma\left(\frac{1}{2\gamma}\right)}\Delta t},
\label{33}
\end{equation}
which corresponds to the radiation era. With relation $\Gamma(1+z)=z\Gamma(z)$, we obtain
\[
\frac{\Gamma\left(\frac{2\gamma+1}{2\gamma}\right)\Gamma\left(\frac{1-\gamma}{2\gamma}\right)}
{\Gamma\left(\frac{\gamma+1}{2\gamma}\right)\Gamma\left(\frac{1}{2\gamma}\right)}=\frac{1}{1-\gamma}.
\]
Taking the duration of the inflation to be $\triangle t \approx 10^{-32}s$ and the values of $\kappa=\sqrt{8\pi G}$ and $a(t_{end})= (2\gamma-1)^{1/(4\gamma)} \beta^{1/4}\sqrt{B_0}$, one can find from Eq. (33) the coupling $\beta$
\begin{equation}
\beta=\frac{2\kappa^2\Delta t^2}{3(2\gamma-1)^{1/\gamma} }.
\label{34}
\end{equation}
\begin{figure}[h]
\includegraphics[height=4.0in,width=4.0in]{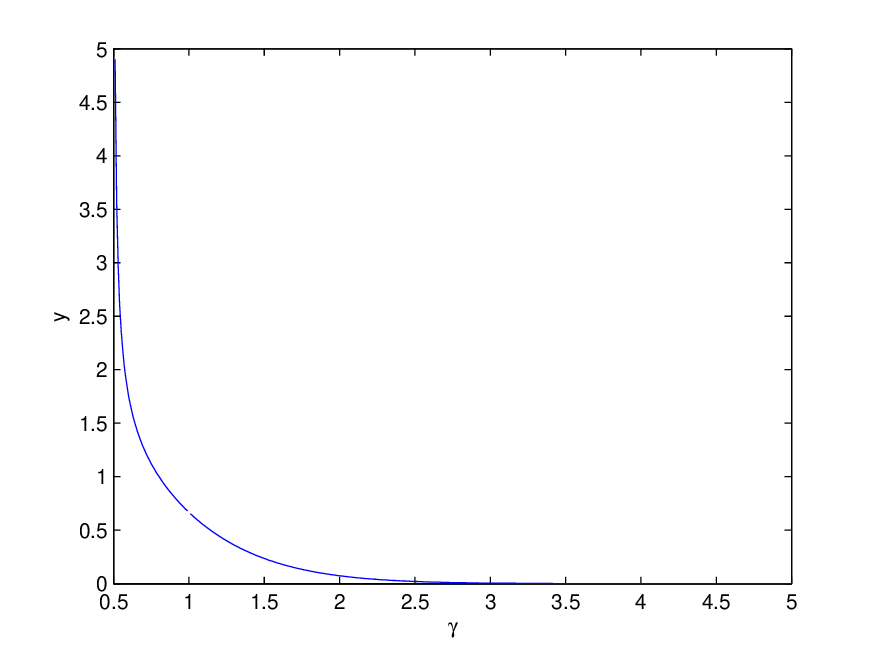}
\caption{\label{fig.6}The function $y=\beta/(\kappa^2\Delta t^2)$ vs. $\gamma$.}
\end{figure}
Figure 6 shows the dependance of $y=\beta/(\kappa^2\Delta t^2)$ versus $\gamma$, where we use $a(t_{end})= (2\gamma-1)^{1/(4\gamma)}\beta^{1/4}\sqrt{B_0}$. According to Fig. 6 when parameter $\gamma$ increases the coupling $\beta$ decreases to have the same time of the universe inflation.


\subsection{Sound speed, causality and unitarity}

The causality holds when the sound speed is less than the local speed of light ($c_s\leq 1$) \cite{Quiros}. If the square sound speed is positive ($c^2_s> 0$) a classical stability takes place. From Eq. (10) one obtains the speed squared of sound
\begin{equation}
c^2_s=\frac{dp}{d\rho}=\frac{dp/d B}{d\rho/dB}=\frac{(4\gamma-1)(\gamma-1)(\beta B^2)^{2\gamma}-(4\gamma^2+5\gamma-2)(\beta B^2)^{\gamma}+1}{3(1+(\beta B^2)^{\gamma})(1-(\gamma-1)(\beta B^2)^{\gamma})}.
\label{35}
\end{equation}
The plots of the function $c_s^2$ vs. $\beta B^2$ are depicted in Fig. 7.
\begin{figure}[h]
\includegraphics[height=4.0in,width=4.0in]{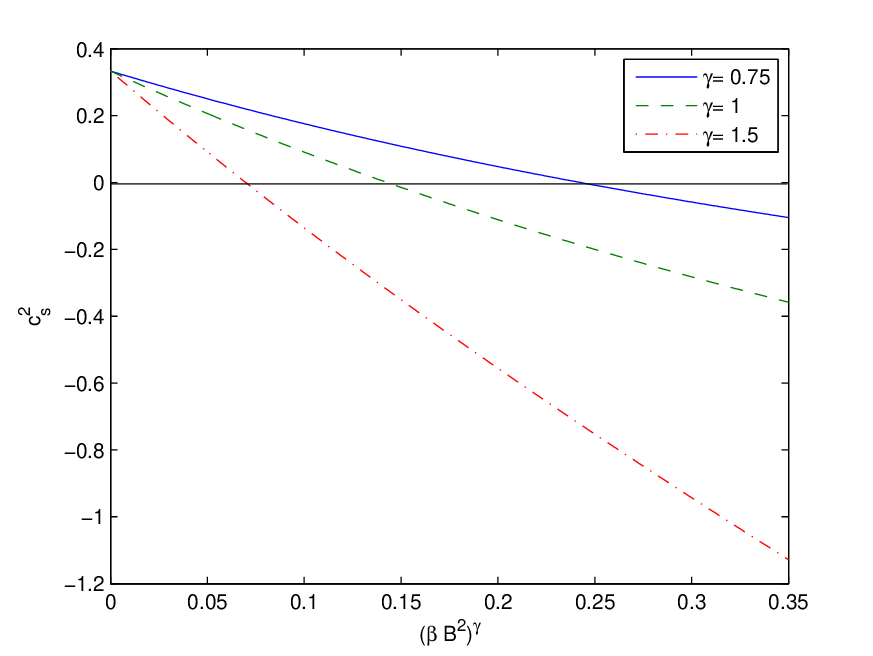}
\caption{\label{fig.7}The function $c_s^2$ vs. $(\beta B^2)^{\gamma}$.}
\end{figure}
Figure 7 shows that at increasing parameter $\gamma$  the causality and unitarity take place for smaller background magnetic field.
It follows from Eq. (35) that $c^2_s=0$  at $\gamma=3/4$, $\beta B^2=(\sqrt{18}-4)^{4/3}\approx0.15$;  $\gamma=1$, $\beta B^2=1/7\approx 0.14$;  $\gamma=1.5$, $\beta B^2=(2.9-\sqrt{8.01})^{2/3}\approx 0.17$.
Making use of Eq. (12) we obtain that acceleration occurs at $\gamma=3/4$, $\beta B^2>0.5^{-4/3}\approx 2.5$;  $\gamma=1$, $\beta B^2>1$;  $\gamma=1.5$, $\beta B^2>2^{-2/3}\approx 0.63$. Thus, at the acceleration phase, the classical stability, causality and unitarity are broken.

\section{Cosmological parameters}

By virtue of Eqs. (10) one obtains
\begin{equation}
p=-\rho+\frac{2 B^2)[1-(\gamma-1)(\beta B^2)^\gamma]}{3[(\beta B^2)^\gamma+1]^2},
\label{36}
\end{equation}
\begin{equation}\label{37}
 2\beta\rho [(\beta B^2)^\gamma+1]-\beta B^2=0.
\end{equation}
By solving Eq. (37) for a particular $\gamma$, one can express $\beta B^2$ as a function of $\rho$. Then Eq. (36) may be represented as
EoS for the perfect fluid
\begin{equation}
p=-\rho+f(\rho),~~f(\rho)=\frac{2z[1-(\gamma-1)z^\gamma]}{3\beta[z^\gamma+1]^2},
\label{38}
\end{equation}
where $z=\beta B^2$ is the solution to Eq. (37) ($2\beta\rho [z^\gamma+1]-z=0$) and $z$ is a function of $\rho$. Analytical solution to Eq. (37) is possible only for several values of $\gamma$. The case $\gamma=1$ was studied in \cite{Kruglov0}. Here, we will investigate the case $\gamma=2$. For this case the solution to Eq. (37) is given by
\begin{equation}\label{39}
z=\frac{1- \sqrt{1-16(\beta\rho)^2}}{4\beta\rho}.
\end{equation}
In the following, we use the negative branch of root squared to have the positive physical function $f(\rho)$ as $p+\rho>0$ (see Eq. (14)). Then from Eqs. (38) and (39) we obtain (at $\gamma=2$)
\begin{equation}\label{40}
f(\rho)=\frac{4\rho}{3}\sqrt{1-16(\beta\rho)^2}.
\end{equation}
In the case when $|f(\rho)/\rho|\ll 1$, the expressions for the spectral index $n_s$, the tensor-to-scalar ratio $r$, and the running of the spectral index $\alpha_s=dn_s/d\ln k$ are given by \cite{Odintsov}
\begin{equation}
n_s\approx 1-6\frac{f(\rho)}{\rho},~~~r\approx 24\frac{f(\rho)}{\rho},~~~\alpha_s\approx -9\left(\frac{f(\rho)}{\rho}\right)^2.
\label{41}
\end{equation}
From Eq.(41) we obtain the relations
\begin{equation}
r=4(1-n_s)=8\sqrt{-\alpha_s}=32\sqrt{1-16(\beta\rho)^2}.
\label{42}
\end{equation}
The PLANCK experiment \cite{Ade} and WMAP data \cite{Komatsu}, \cite{Hinshaw} gave the result
\[
n_s=0.9603\pm 0.0073 ~(68\% CL),~~~r<0.11 ~(95\%CL),
\]
\begin{equation}
\alpha_s=-0.0134\pm0.0090 ~(68\% CL).
\label{43}
\end{equation}
From the requirements $|f(\rho)/\rho|\ll 1$  and $16(\beta\rho)^2<1$ one finds
\begin{equation}
\frac{1}{4}>\beta\rho\gg\frac{\sqrt{7}}{16}.
\label{44}
\end{equation}
Making use of Eqs. (42) at $r=0.13$ we obtain the values for the spectral index $n_s=0.9675$ and the running of the spectral index $\alpha_s=-2.64\times 10^{-4}$. By virtue of Eq. (42), one finds the value $\beta\rho\approx 0.249998$ which satisfies Eq. (44) and
corresponds to values of cosmological parameters. To find the value of the magnetic field one can use Eq. (10) (at $\gamma=2$) by fixing the coupling $\beta$ and using the value $\beta\rho\approx 0.249998$. The plot of the function $\beta B^2$ versus $\beta\rho$ is depicted in Fig. 8.
\begin{figure}[h]
\includegraphics[height=4.0in,width=4.0in]{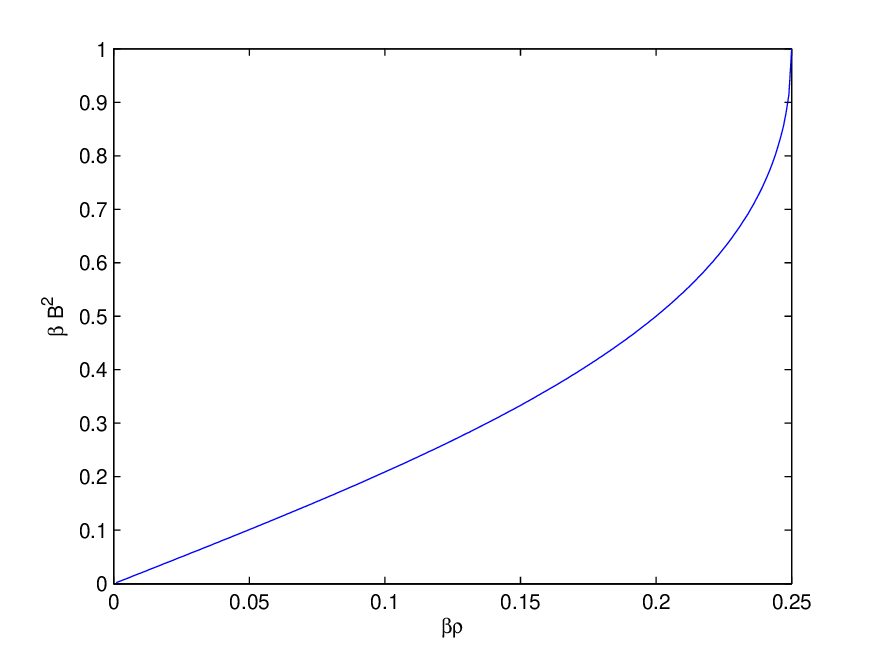}
\caption{\label{fig.8}The function $\beta B^2$ vs. $\beta\rho$.}
\end{figure}
For $\beta\rho\approx 0.249998$ we have $\beta B^2\approx 0.996$.

\section{Summary}

We have studied Einstein's gravity coupled to nonsingular NED with two parameters. The stochastic magnetic fields of background as a source of gravity were implied. The range of magnetic fields, when the inflation of the universe takes place, has been obtained. It was shown that there are not singularities of the energy density and pressure at $\gamma \geq 1$. We have demonstrated that as the scale factor goes to infinity, EoS of ultra-relativistic case takes place. It was shown that singularities of curvature invariants are absent for $\gamma\geq 1$. We have investigated  the evolution of the universe. The dependence of the scale factor on the time has been obtained. To study the evolution of universe, the deceleration parameter as a function of the scale factor has been calculated. To analyse the amount of the inflation we have calculated e-foldings. It was shown that for some parameters the reasonable e-foldings $N\approx70$ holds. The duration of the universe inflation as the function of parameters $\beta$ and $\gamma$ has been obtained.
We computed the speed of sound to study the causality and unitarity principles in our model. It was found the range of background magnetic fields when causality and unitarity hold.
The cosmological parameters, the spectral index $n_s$, the tensor-to-scalar ratio $r$, and the running of the spectral index $\alpha_s$, were computed. We showed that $n_s$, $r$, and $\alpha_s$ are in agreement with the PLANK and WMAP data.

Let us discuss the similarities and differences of our model of the universe inflation with the model in Ref. \cite{Benaoum}.\\
\textbf{Similarities}: The Ricci scalar curvature, the Ricci tensor squared and the Kretschmann scalar have no singularities at early/late stages, as $a\rightarrow 0$. In the model  \cite{Benaoum} and in our model $w =1/3$, $q=1$ as $a\rightarrow \infty$.
The early universe did not have Big-Bang singularity and accelerated in the past. The slow-roll parameters such as spectral index $n_s$, and tensor-to-scalar ratio $r$ were analysed and compared with results of observational data. The classical stability and causality are studied.\\
\textbf{Differences}: The behaviour of EoS parameter $w$ and deceleration parameter $q$ as $a\rightarrow 0$ are different. In our model $w=(1-4\gamma)/3$, $q=1-2\gamma$ but in the model of \cite{Benaoum}, $w=-1$, $q=-1$ for any $\alpha$ as $a\rightarrow 0$.
The duration of the universe inflation as the function of parameters $\beta$ and $\gamma$ was computed in our model. Phase space analysis was performed in [36]. Corrections of NED Lagrangian to Maxwell's Lagrangian are different at weak-field limit.

\end{document}